**AI as a deliberative partner fosters intercultural empathy for Americans but fails for Latin American participants**.

Isabel Villanueva, Tara Bobinac[b], Binwei Yao[b], Junjie Hu[b], Kaiping Chen[a*]
[a]Department of Life Sciences Communication, University of Wisconsin-Madison, Madison, WI, 53706, USA
[b]Department of Computer Sciences, University of Wisconsin-Madison, Madison, WI, 53706, USA

*Corresponding Author: Kaiping Chen

Email: kchen67@wisc.edu

## Abstract

Despite increasing AI chatbot deployment in public discourse, empirical evidence on their capacity to foster intercultural empathy remains limited. Through a randomized experiment, we assessed how different AI deliberation approaches—cross-cultural deliberation (presenting other-culture perspectives), own-culture deliberation (representing participants' own culture), and non-deliberative control—affect intercultural empathy across American and Latin American participants. Cross-cultural deliberation increased intercultural empathy among American participants through positive emotional engagement, but produced no such effects for Latin American participants, who perceived AI responses as culturally inauthentic despite explicit prompting to represent their cultural perspectives. Our analysis of participant-driven feedback, where users directly flagged and explained culturally inappropriate AI responses, revealed systematic gaps in AI's representation of Latin American contexts that persist despite sophisticated prompt engineering. These findings demonstrate that current approaches to AI cultural alignment—including linguistic adaptation and explicit cultural prompting—cannot fully address deeper representational asymmetries in AI systems. Our work advances both deliberation theory and AI alignment research by revealing how the same AI system can simultaneously promote intercultural understanding for one cultural group while failing for another, with critical implications for designing equitable AI systems for cross-cultural democratic discourse.

## Introduction

Public deliberation is at the core of democracy. However, large-scale discussions often deviate from deliberative ideals, becoming dominated by one-sided advocacy, misinformation, and limited exposure to diverse viewpoints (1, 2). Although structured deliberation can mitigate these challenges, it remains resource intensive and inaccessible for everyday conversations (3). Advances in Artificial Intelligence (AI) offer the potential to scale deliberative interactions, yet it remains unclear whether direct conversations with AI conversational partners can achieve key deliberative outcomes such as empathy, perspective taking, and cross-cultural understanding (3, 4). Most research on AI and deliberation examines AI as a moderator facilitating human-to-human dialogues (1, 5, 6). Our study instead investigates AI as a deliberative conversation partner, examining how interactions with an AI programmed to present diverse cultural perspectives influence users' intercultural empathy. This approach offers two advantages: it reflects how AI typically functions in real-world contexts where interactions occur one-on-one; and it explores scaling deliberative experiences beyond formal settings, making perspective-enhancing conversations accessible without group coordination constraints. This addresses the growing prevalence of individual AI interactions and their deliberative potential.

Deliberation requires a reciprocal discourse that emphasizes rationality, intentional reasoning, and reflection—where participants engage in reasoned exchange of viewpoints to reach better understanding (4). Intercultural empathy—the ability to understand and emotionally connect with perspectives from different cultural backgrounds—represents a crucial deliberative outcome, as democratic discourse increasingly requires citizens to engage constructively across cultural divides. Although Large Language Models (LLMs) lack deliberative intentionality, they could have the potential to facilitate user reflection and foster intercultural learning, as people often apply social rules between human-to-human communication to computer-mediated contexts through treating computers as social actors (7). Recent work on conversations with AI has also shown to improve users' reciprocity and listening in discourse (3).

However, realizing this deliberative potential depends critically on whether AI can authentically engage users from diverse cultural backgrounds. A critical challenge for AI-assisted deliberation is cultural alignment—whether users perceive AI responses as culturally appropriate and reflective of their experiences. Research found that LLMs align more closely with Western, White, and educated perspectives (8–10), raising concerns about engaging diverse cultural groups. These previous studies have focused on benchmarking static datasets rather than examining how misalignment manifests during real-time interactions, a gap our study addresses. AI systems may encode deeper cultural and cognitive patterns, with research showing that LLMs exhibit different social orientations and cognitive styles when operating in different languages—displaying more interdependent social orientations and holistic cognitive styles in Chinese versus more independent orientations and analytic styles in English (11). Our study empirically examines this issue by comparing US and Latin American participants—a theoretically important contrast given that Latin American cultures often emphasize more collectivist values and relationship-centered communication styles, potentially creating greater misalignment with AI systems primarily trained on Western discourse patterns.

This cultural alignment challenge raises important theoretical questions about how AI systems might foster intercultural empathy. Contact theory suggests that exposure to different cultural perspectives can reduce prejudice and increase empathy by providing opportunities to understand other viewpoints (12, 13). From this perspective, an AI system that exposes users to cross-cultural perspectives—even if imperfectly represented—might still promote intercultural understanding by broadening users' awareness of how others think about contentious issues.

However, authenticity in intercultural communication emphasizes that meaningful cross-cultural engagement depends on genuine, culturally accurate representation (14, 15). If AI systems cannot authentically represent cultural perspectives—particularly users' own cultural contexts—they may lack the credibility necessary to facilitate effective deliberation. This raises questions about whether current AI systems possess sufficient cultural knowledge to serve as authentic deliberative partners, especially for non-Western users.

These theoretical frameworks suggest examining AI deliberation through two complementary lenses: testing whether AI can promote intercultural empathy through cross-cultural exposure (contact-based approach), while simultaneously evaluating whether AI can authentically represent cultural perspectives well enough to be perceived as credible deliberative partners (authenticity-based approach). We propose the following research questions:

**RQ1. To what extent does direct deliberative interaction with an AI chatbot presenting cross-cultural perspectives foster intercultural empathy and perspective-taking, and how does this effect vary between US and Latin American participants?**

**RQ2. How do US and Latin American participants evaluate the cultural accuracy of AI responses when the AI is prompted to represent their own cultural perspectives during deliberative conversations?**

Our study makes three key contributions. Theoretically, we advance deliberation theory by examining direct human-AI deliberative exchanges rather than AI-facilitated group discussions, demonstrating how AI conversational partners might influence intercultural empathy development through both cross-cultural exposure and cultural authenticity mechanisms. Methodologically, we introduce an innovative multi-method approach that combines controlled experimental conditions with real-time emotional feedback and participant-driven identification of culturally inappropriate AI responses—moving beyond static benchmarking to capture dynamic cultural misalignment as it unfolds during actual conversations. Empirically, we provide the first systematic investigation of LLMs as cross-cultural deliberative partners, revealing significant cultural knowledge asymmetries that persist despite sophisticated prompt engineering. Our findings demonstrate that the same AI system can simultaneously promote intercultural empathy for one cultural group while failing to do so for another, with implications for designing AI systems that facilitate equitable cross-cultural dialogue in democratic societies.

## Materials and Methods

### Experimental Design

We conducted an online dialogue experiment with 3 conditions (own-culture deliberation, cross-cultural deliberation, and non-deliberative control) across two demographic groups (American vs. Latin American participants) to examine whether exposure to cross-cultural deliberative chatbots was associated with increases in intercultural empathy and learning. Further, we investigated to what extent Latin American participants perceived cultural alignment when exposed to the own-culture deliberation condition, compared to U.S. participants.

We wrote programming scripts to instruct Llama, an open-sourced LLM, to use two types of system prompts (the initial instructions that define the AI's role and behavior): one acted as a deliberation partner representing diverse perspectives within each national context (either the US or a specific Latin American country: Brazil, Colombia, Mexico, or Nicaragua), specifically guiding the LLM to present pros and cons that different populations within that country hold regarding high-stake issues (e.g., abortion, LGBTQ+, death penalty). This approach acknowledges the cultural heterogeneity within national boundaries and avoids treating any country as having a singular cultural perspective. The non-deliberator prompt type instructed the LLM to be a stubborn conversational partner that does not introduce any culturally relevant background information and instead adheres to a one-sided opinion of an issue *(for details of User Interface and System Prompts, see Online Supplemental Material S4)*. To capture the effects of these different AI interactions, we employed multiple measurement approaches.

**Survey and Measures**

Participants (*N*=600; U.S. = 315, Latin America = 285) first completed a pre-chat survey which included a battery of questions measuring intercultural empathy (17) and participants' attitudes of the to-be-discussed issues (e.g., GMOs, abortion). See **Figure 1** about our procedures for the experiment.

Intercultural empathy was measured using a 12-item scale, which included items such as "In cross-cultural interactions <u>about [topic]</u>, I notice how our cultures are similar to and different from one another" and "I can readily take the perspective of someone from another culture <u>when thinking about [topic]</u>." Respondents reported their response to these items on a scale from 1 (strongly disagree) to 6 (strongly agree). Then, attitudes toward each topic were measured using a single item from the World Value Survey (WVS; Haerpfer et al., 2022) that asked respondents to rate whether each topic "can always be justified, never be justified, or something in between", using a scale from 1 (never justified) to 10 (always justified).

Participants were then directed to a User Interface that allowed them to have 10-15 rounds of real-time conversation with a randomly assigned Llama bot. During each round, participants rated the chatbot's responses using emojis which were used as a measure of emotional response to the chatbot. At the end of the dialogue, participants also provided feedback on the chatbot's responses using tags such as "culturally (un)knowledgeable", "(dis)respectful", and "(not) open to listening to user", and were required to provide written comments about at least three of the chatbot's responses which were later analyzed for cultural alignment. After completing the dialogue with the bot, participants were redirected to a post-chat survey remeasuring intercultural empathy and asking about users' satisfaction with the bot.

Satisfaction with the bot was measured with a five-item index, which included items such as "My interaction with the chatbot was satisfying" and "Chatting with the chatbot was enjoyable." Participants reported their agreement with these statements on a five-point scale, 1 (strongly disagree) to 5 (strongly agree). Finally, participants were asked an open-ended question to report what they had learned from the interaction: "Based on your conversation with the chatbot, what have you learned about how people from [country] perceive [topic]?".

It is important to note that participants chatted with Llama and completed the surveys in their native language (i.e., English if they were from the United States, Spanish if they were from Mexico, Colombia, or Nicaragua, or in Portuguese if they were from Brazil).

**Topic Selection**

The topics discussed during chatbot interactions were selected from a set of commonly addressed topics in the World Value Survey (WVS; Haerpfer et al., 2022). Specifically, the WVS assesses attitudes towards various topics by asking respondents to rate whether each topic "can always be justified, never be justified, or something in between", using a scale from 1 (never justified) to 10 (always justified). Among these surveyed issues are abortion, the death penalty, and homosexuality. For this study, we chose topics that were politically relevant across the various countries, but exhibited notably different average views between Latin American and the United States. Descriptive statistics and independent t-test results are presented in the supplemental material to compare the differences in viewpoints between Latin America and the United States.

Additionally, we chose a scientific topic - genetically modified organisms (GMOs) – which is not covered in the WVS. GMOs remain a contentious topic in countries around the world (Pew Research Center, 2020). In Latin America, public opinion surrounding GMOs varies widely by country, with countries such as Brazil facing less opposition due to its prominent leadership role in the implementation of the technology (Sikora et al., 2021). Meanwhile, countries such as Mexico remain generally opposed, driven by factors such as the preservation of native crops, safeguarding cultural heritage, and fear surrounding loss of biodiversity (Woźniak-Gientka et al., 2022). We chose this scientific topic because public opinion toward it is deeply influenced by cultural considerations.

**Analysis**

To examine the relationship between experimental conditions (cross-culture deliberation vs. control), changes in intercultural empathy, and potential mediators, we conducted two mediation analyses. The first analysis explored whether emotional activation mediated the relationship between experimental conditions and changes in intercultural empathy from the presurvey to the postsurvey. The second analysis tested whether satisfaction with the chatbot served as the mediator.

We also qualitatively analyzed participants' learning outcomes across the cross-cultural deliberation and control conditions using grounded theory—an iterative qualitative analysis method that develops themes from data (18). Open-ended responses were coded to identify common themes and phrases and categorized based on whether participants reported learning about: (1) cultural aspects of the topic, (2) facts about the topic, (3) others' perspectives without a reference to culture, or (4) if they stated they learned something without providing examples. A chi-square test was then used to assess whether there were significant differences in reported learning outcomes between Latin American and American participants, or between those in the cross-cultural deliberation and control groups.

To assess perceptions of cultural alignment, we analyzed how participants in their own-culture deliberation condition and the control condition categorized their open-ended comments about the chatbot's responses. The tag categories included: "(dis)respectful", "culturally (un)knowledgeable", "(not) open to listening to the user", and "factually (in)correct". We conducted two-sample t-tests for each response category to determine whether Latin American and American participants differed significantly in their feedback.

We also analyzed the open-ended feedback specific to the own-culture deliberation to further investigate cultural alignment. The negative, open-ended responses of our participants in the own-culture deliberation condition were iteratively coded for common topics, which were then grouped into larger themes, including: criticism of the bot, absence of cultural perspectives, lack of examples, lack of actionable steps, and general dissatisfaction. We then conducted a chi-square test to determine whether the negative feedback themes differed between Latin American and American participants.

## Results

We examined how different types of AI deliberation affect intercultural empathy across cultural groups using a three-condition experiment: cross-cultural deliberation (AI presented other-culture perspectives), own-culture deliberation (AI represented participant's own culture), and non-deliberative control (AI provided one-sided opinions). Our results revealed a striking asymmetry in how AI deliberation affects intercultural empathy and learning: cross-cultural AI exposure fostered positive deliberative outcomes for American participants but not for Latin American participants. This differential effectiveness appears to stem from fundamental differences in how these groups perceived the cultural authenticity of AI responses.

### Cross-Cultural AI Exposure: Divergent Outcomes by Cultural Group

Our central finding reveals that exposure to the cross-cultural deliberation chatbot produced dramatically different outcomes depending on participants' cultural background. While American participants experienced significant increases in intercultural empathy and cultural learning when exposed to cross-cultural AI deliberation, Latin American participants showed no such benefits from the identical intervention.

#### *American participants: Contact Theory in Action*

For American participants, exposure to a cross-cultural deliberative chatbot was indirectly associated with increased intercultural empathy among American participants, mediated by positive emotions and satisfaction with the bot (**Table 1, Figure 2**). Participants in the cross-culture deliberation condition reported an average of 0.35 more positive emotions (on a scale from 1 to 5, or a 7% increase) through emojis during their conversation with the chatbot, which in turn enhanced their intercultural empathy (b = 0.038, SE = 0.018, 95% CI = [0.003, 0.073]). Similarly, those who reported the chatbot as 1 point more satisfying in the post-survey (out of 5) reported, on average, 0.132 points (out of 5, or a 3% increase) higher intercultural empathy than they did in the beginning of the survey (b = 0.053, SE = 0.021, 95% CI = [0.012, 0.94]).

**Learning outcomes further demonstrated this effectiveness.** When asked in the post-chat survey what they had learned about the topic from their chatbot conversation, American participants showed significant differences between conditions. Among those exposed to cross-cultural AI, 39 (16.5%) reported learning cultural information about the topic, compared to none (0%) in the control group, $\chi^2$ (5, $N$ = 236) = 15.03, $p$ = .010. For instance, one American reported "*I've learned that the people of Mexico accept the usage of GMOs, but they want to use them responsibly in order to make sure that the earth is taken care of and that small farmers aren't squeezed out of the marketplace through their usage by huge corporations*."

### *Latin American Participants: When Cross-Cultural Exposure Fails*

In contrast, for Latin American participants, the cross-cultural deliberative condition was not significantly associated with increased intercultural empathy, even when considering indirect pathways through emotion (b = 0.013, SE = 0.012, 95% CI = [-0.010, 0.036]) or bot satisfaction (b = -0.001, SE = 0.001, 95% CI = [-0.020, 0.08]).

**Learning outcomes mirrored this pattern.** When asked to write down what they have learnt about the discussed topic with the chatbot in the post-survey, Latin American participants showed minimal differences between those in the cross-cultural condition and those in the control condition, $\chi^2$ (5, $N$ = 173) = 5.17, $p$ = .39. Only 2 (1%) participants in the control group and 5 (3%) in the cross-cultural condition reported learning cultural information about the topic. Similarly, 33 (19%) participants in the control group and 39 (23%) participants in the cross-cultural group reported learning about others' perspectives. One Latin American participant stated about the death penalty: "*Some consider it an effective deterrent to crime, while others worry about miscarriages of justice, discrimination, and its impact on morality. Furthermore, cultural, religious, and media influences play a significant role in how people perceive this issue [translated from Spanish]*." A full breakdown of the different types of learning outcomes comparing American participants vs. Latin American participants can be found in **Figure 3**.

### Explaining the Divergence: Cultural Authenticity Matters

The key to understanding this asymmetry lies in how participants perceived the cultural authenticity of AI responses when the chatbot was prompted to represent their own culture. American participants in the own-culture deliberative chatbot condition categorized the chatbot's responses as "respectful" 70% more often (M = 3.72, SD = 4.40) than Latin American participants (M = 2.18, SD = 3.47), t(83.24) = 2.14, $p$ = .038. In contrast, Latin American participants tagged the chatbot's responses as "factually incorrect" over twice as often (M = 1.20, SD = 1.76) than their American counterparts in the own-culture deliberation condition (M= 0.50, SD = 1.20), t(132.07) = -2.79, $p$ = 0.006.

These results suggest that while Americans found AI representations of their culture relatively authentic, Latin American participants perceived significant cultural misalignment even when the AI was explicitly prompted to represent their perspectives.

### Evidence from User Evaluations: Americans vs. Latin Americans

The depth of this cultural misalignment becomes clear through participants' detailed feedback collected after their conversations. After finishing their dialogue with the chatbot, participants were directed to a feedback page where they tagged and commented on at least three chatbot responses. Our analysis of 983 written comments from participants in the own-culture condition revealed striking differences: only 6 (0.61%) of American participants' 416 comments were negative, while 87 (8.85%) of Latin American participants' 567 comments were negative, $\chi^2$ (1, $N = 983$) = 52.53, $p < 0.001$. See **Figure 4** for a comparison of feedback patterns between American and Latin American participants.

**The content of critiques differed fundamentally by cultural group.** American participants' negative comments focused on criticizing the chatbot's style (e.g., the response was too repetitive), while Latin American participants emphasized how the bot was lacking cultural perspectives. For instance, one Mexican participant stated, "*GMOs pose a dilemma in Mexico due to cultural, environmental, and economic factors. In Mexico, we have a rich and diverse culture that values agricultural tradition, which could be affected by GMOs in terms of biodiversity and genetic heritage. However, we also face significant economic challenges such as hunger and rural poverty…"* This detailed elaboration suggests participants felt compelled to provide the cultural context and nuanced perspective that the AI failed to deliver.

Together, these findings reveal that AI's effectiveness as a deliberative partner depends critically on perceived cultural authenticity. While cross-cultural exposure through AI can foster intercultural empathy and learning, this contact-based mechanism appears to require that users perceive the AI source as culturally credible. For Latin American participants, cultural misalignment with AI responses appears to have created an authenticity barrier that prevented the positive emotional engagement necessary for intercultural learning.

## Discussion

Our findings reveal a striking asymmetry in AI's capacity to foster intercultural empathy: while American participants experienced increased empathy and learning through cross-cultural AI deliberation, Latin American participants showed no such benefits and perceived the same AI responses as culturally inauthentic. This differential effectiveness challenges assumptions about AI's universal capacity to serve as deliberative partners and highlights critical limitations in current AI systems' cultural representation.

The asymmetric learning outcomes further underscore these disparities. American participants exposed to the cross-cultural AI deliberation were significantly more likely to report learning about cultural perspectives (16.5% vs. 0% in control), with participants noting insights such as learning that people of Mexico accept the usage of GMOs, but they want to use them responsibly. In contrast, Latin American participants showed minimal differences in learning outcomes between the cross-culture deliberation condition and the control condition, suggesting that the AI's cultural presentations failed to provide meaningful new insights for these users.

### Theoretical Implications for Contact Theory and Authenticity Theory

These findings provide nuanced support for both theoretical frameworks established in our introduction, while revealing important boundary conditions. Our results suggest that contact theory (12, 13) may extend to AI-mediated contexts for American participants—exposure to cross-cultural AI perspectives successfully increased intercultural empathy and cultural learning. However, this AI-mediated contact mechanism appears to be contingent on participants' perception of the AI source as credible and culturally competent, suggesting that the conditions necessary for effective intergroup contact may be even more stringent in computer-mediated contexts.

For Latin American participants, the failure of cross-cultural exposure to increase empathy suggests that authenticity concerns (14, 15) may override contact benefits. When participants perceive AI responses as culturally inauthentic—as evidenced by their tagging of responses as "culturally unknowledgeable" and "factually incorrect"—the contact mechanism appears to break down. This suggests that authenticity may be a prerequisite for effective intergroup contact in AI-mediated contexts, particularly for users from underrepresented cultural groups who may be more sensitive to misrepresentation.

Importantly, our own-culture deliberation condition results further illuminate authenticity theory. American participants found AI representations of their own culture relatively authentic (tagging responses as "respectful" 70% more often than Latin Americans), while Latin American participants frequently perceived cultural misrepresentation. This pattern suggests that current AI systems possess sufficient cultural knowledge to achieve baseline authenticity for American users but lack the nuanced understanding necessary for authentic representation of Latin American perspectives.

### Underlying Technical Factors

Our findings highlight that using one's native language in deliberative conversations with AI does not eliminate cultural biases inherent in LLMs. Despite engaging in conversations in Spanish or Portuguese with an LLM specifically instructed to adopt the persona of a cultural deliberator from their respective countries, Latin American participants were more likely than US participants to perceive AI responses as culturally unknowledgeable. A central factor contributing to this issue could be the pronounced imbalance in multilingual training data. Given that only a small proportion (approximately 5%) of LLM training data is non-English (16), linguistic and cultural asymmetries persist, influencing AI interactions across cultural contexts. While training data imbalances help explain these disparities, our methodological approach revealed how these limitations manifest in practice.

### Methodological Contributions

Prior studies assessed AI alignment primarily through analytical benchmarking approaches (8, 9), where researchers evaluate AI outputs against predefined criteria or survey responses. Our research extends beyond such evaluation frameworks by examining how cultural misalignment manifests during real-time conversations. This methodology captures users' direct, in-the-moment reactions to perceived cultural inaccuracies in AI responses, revealing how alignment issues emerge through successive conversational turns rather than through post-hoc analysis of isolated outputs.

**Mechanisms of Cultural Misalignment: Insights from Qualitative Feedback**

The qualitative feedback from participants provides valuable explanations for the observed cultural misalignment. Our analysis of 983 written comments revealed stark differences in both the frequency and nature of negative feedback: only 0.61% of American participants' comments were negative (primarily focusing on stylistic issues like repetitiveness), while 8.85% of Latin American participants' comments were negative and concentrated on cultural inadequacy.

The content of these critiques further illuminates the different user experiences. While American participants focused on surface-level stylistic concerns, Latin American participants provided substantive critiques about missing cultural depth. For instance, one Mexican participant felt compelled to elaborate extensively on the cultural, environmental, and economic complexities surrounding GMOs in Mexico, emphasizing agricultural traditions, biodiversity concerns, and rural poverty challenges that the AI had failed to address. These detailed feedback suggest that the participant felt compelled to provide the cultural context and nuanced perspective that the AI failed to deliver. These qualitative insights point to a deeper mechanism underlying the observed cultural disparities: the central role of emotional engagement in AI deliberation.

**The Central Role of Affective Mechanism in AI-mediated Deliberation**

While deliberative democracy traditionally emphasizes reciprocity, reason-giving, and perspective-taking (4), our findings suggest the impact of deliberation on perspective-taking operates through the affective mechanism. Our findings reveal that emotional engagement serves as a crucial mediator in AI-facilitated intercultural empathy development, challenging traditional rational-focused models of deliberation. For American participants, positive emotional responses during conversations (measured through emoji feedback) directly predicted increases in intercultural empathy ($b = 0.038$, $SE = 0.018$, 95% CI = [0.003, 0.073]). Similarly, participants who found their interactions with the AI satisfying showed corresponding increases in empathy ($b = 0.053$, $SE = 0.021$, 95% CI = [0.012, 0.94]).

Importantly, this affective pathway was absent for Latin American participants, where neither positive emotions ($b = 0.013$, $SE = 0.012$, 95% CI = [-0.010, 0.036]) nor satisfaction ($b = -0.001$, $SE = 0.001$, 95% CI = [-0.020, 0.08]) mediated empathy outcomes. This suggests that affective engagement may be contingent on perceived authenticity—when users perceive AI as culturally misaligned, the emotional connection necessary for empathy development fails to occur.

This pattern offers significant theoretical implications for understanding AI's role in deliberative processes. Unlike human-to-human deliberation where cognitive argumentation may suffice, AI-mediated deliberation appears to require an additional layer of affective credibility. Users must not only find AI arguments logical but must also experience positive emotional engagement with the AI partner itself. For Latin American participants, cultural misalignment appeared to create an affective barrier that prevented the emotional engagement necessary for intercultural learning, regardless of the AI's argumentative content. Our finding aligns with Young's (12) point that effective deliberation requires acknowledging both cognitive and affective components

Ultimately, our findings demonstrate that neither linguistic alignment nor explicit system prompting to adopt culturally-specific deliberative roles mitigates deeper cultural knowledge gaps within AI conversational partners. By highlighting this limitation, we extend both deliberation theory and AI research, emphasizing the challenge of designing systems that genuinely engage diverse populations in intercultural deliberation.

**Practical Implications for AI System Design**

Our findings also provide immediate implications for developers and policymakers working on AI systems intended for cross-cultural contexts. First, training data diversification alone is insufficient—our results suggest that even when AI systems operate in users' native languages and are explicitly prompted to represent local cultural perspectives, fundamental knowledge gaps persist. This indicates that current approaches to cultural alignment, which focus primarily on multilingual capabilities and surface-level cultural prompting, need substantial revision.

Second, evaluation frameworks must incorporate user-driven assessments rather than relying solely on expert benchmarking. Our participant-driven tagging approach revealed cultural misalignments that would likely be missed by traditional evaluation methods. AI developers should implement real-time feedback mechanisms that allow users from different cultural backgrounds to identify and flag culturally inappropriate responses during system development and deployment.

Third, affective engagement should be treated as a prerequisite for effective AI deliberation, not a byproduct. Our findings suggest that AI systems must achieve both cognitive accuracy and emotional credibility to facilitate intercultural learning. This requires developing AI that can engage users in culturally resonant ways, potentially through more sophisticated understanding of cultural communication styles, values, and priorities.

Finally, deployment strategies should acknowledge differential effectiveness across cultural groups. Rather than assuming universal applicability, AI deliberation systems should be pilot-tested extensively with diverse user populations and potentially offer different interaction modes for different cultural contexts. The alternative—deploying systems that systematically advantage some cultural groups over others—risks exacerbating rather than addressing intercultural divides in democratic discourse.

**Limitations**

Several limitations should be acknowledged when interpreting these findings. First, our sample focused on participants from the United States and four Latin American countries (Brazil, Colombia, Mexico, Nicaragua), which may limit generalizability to other cultural contexts or regions. Second, our study employed a single LLM (Llama). Results may vary with different AI models. Third, the topics discussed (GMOs, abortion, death penalty, LGBTQ+ issues) were selected for their cross-cultural relevance but may not represent all domains where AI deliberation might be applied. Fourth, while our participant-driven tagging approach captures real-time cultural misalignment, it relies on participants' own cultural awareness and ability to articulate misrepresentation, which may vary across individuals. Finally, our study examined relatively short conversational interactions (10-15 rounds); longer-term engagement patterns with AI deliberative partners remain unexplored.

**Broader Implications and Future Directions**

Ultimately, our findings demonstrate that neither linguistic alignment nor explicit system prompting to adopt culturally-specific deliberative roles mitigates deeper cultural knowledge gaps within AI conversational partners. By highlighting this limitation, we extend both deliberation theory and AI research, emphasizing the challenge of designing systems that genuinely engage diverse populations in intercultural deliberation. A pivotal question is whether AI can function as an authentic deliberative partner or if structural biases confine it to procedural facilitation. Resolving this requires examining AI's epistemic role—its perceived credibility and authority—and its capacity to facilitate authentic intercultural dialogue in democratic societies.

## Figures and Tables

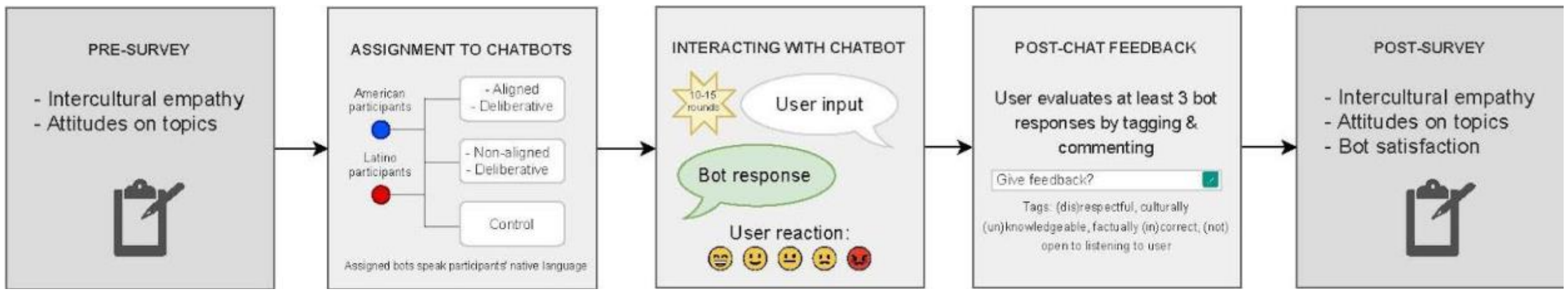

**Figure 1.** Our Online Dialogue Experiment Procedures

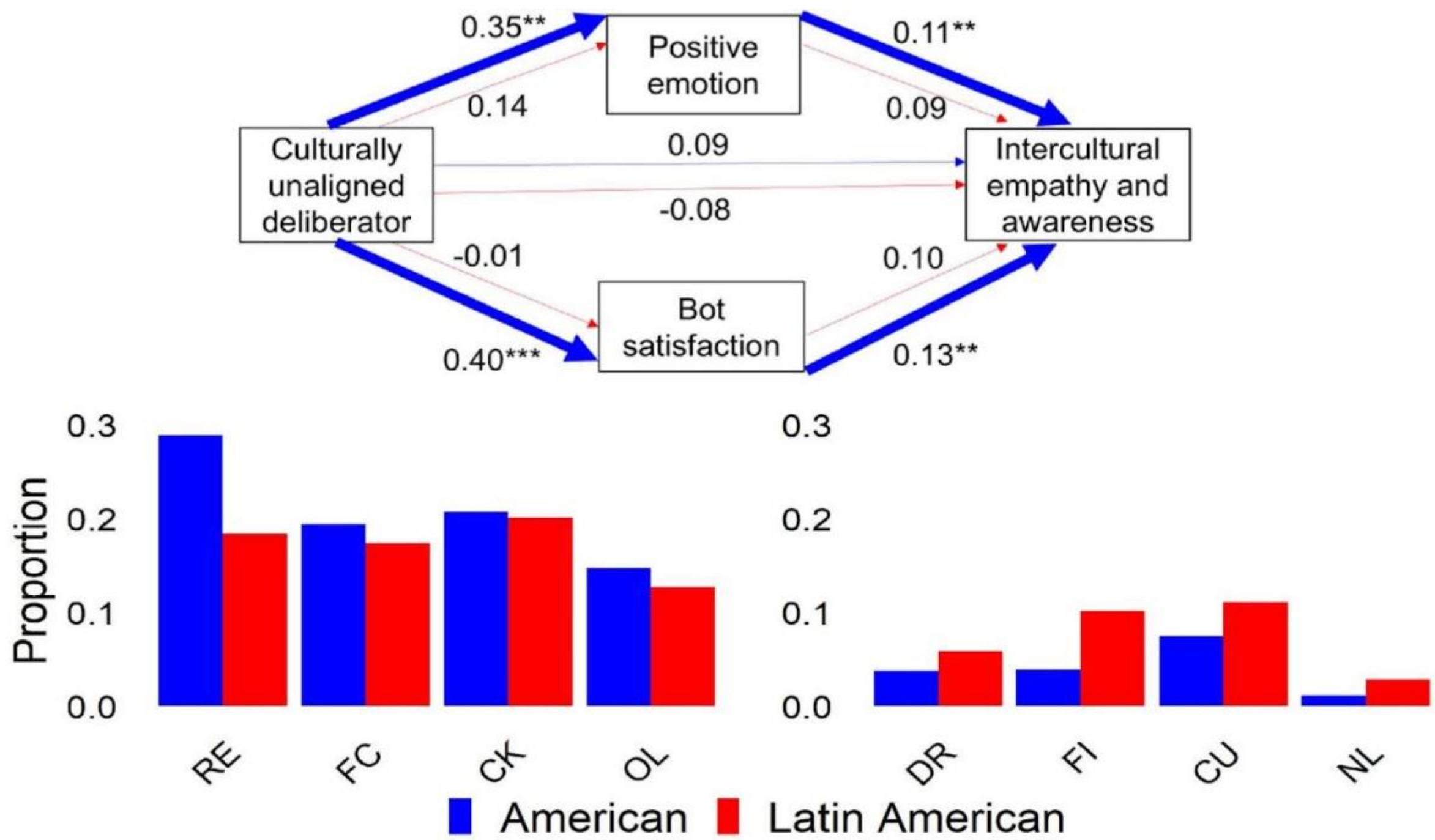

**Figure 2.** Deliberation, Intercultural Empathy, and Cultural Alignment.

Note: Numbers in the top path diagram are standardized regression coefficients. *p < 0.05, **p < 0.01, ***p < 0.001. For the full regression tables of our path analysis results, see Online Supplemental Material S5 and S6. Bottom figure = proportion of each tag used by individuals in the "culturally aligned" condition under which a participant had dialogues with a Llama chatbot that was system prompted to play the role of a deliberative partner of where the participant came from. RE refers to the tag "Respectful", FC refers to "Factually Correct", CK refers to "Culturally Knowledgeable", OL refers to "Open to Listening", DR refers to "Disrespectful", FI refers to "Factually Incorrect", CU refers to "Culturally Unknowledgeable", and NL refers to "Not Open to Listening". For the full regression tables on tag analysis, see Online Supplemental Material S7 and S7.

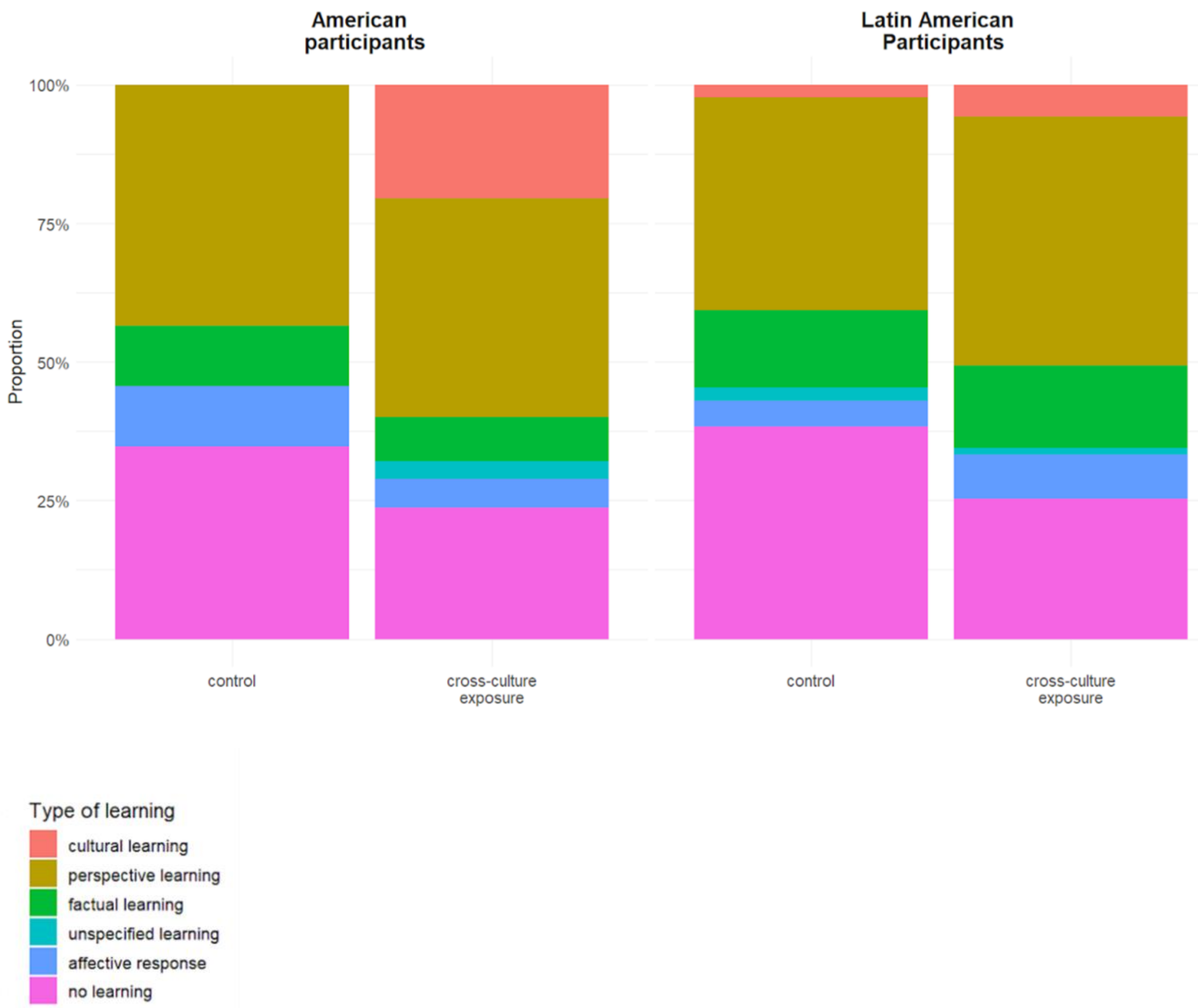

**Figure 3. What U.S. and Latin American participants in the cross-cultural condition wrote in their responses to the open-ended question in the post-survey that asked what they learned from the dialogue**

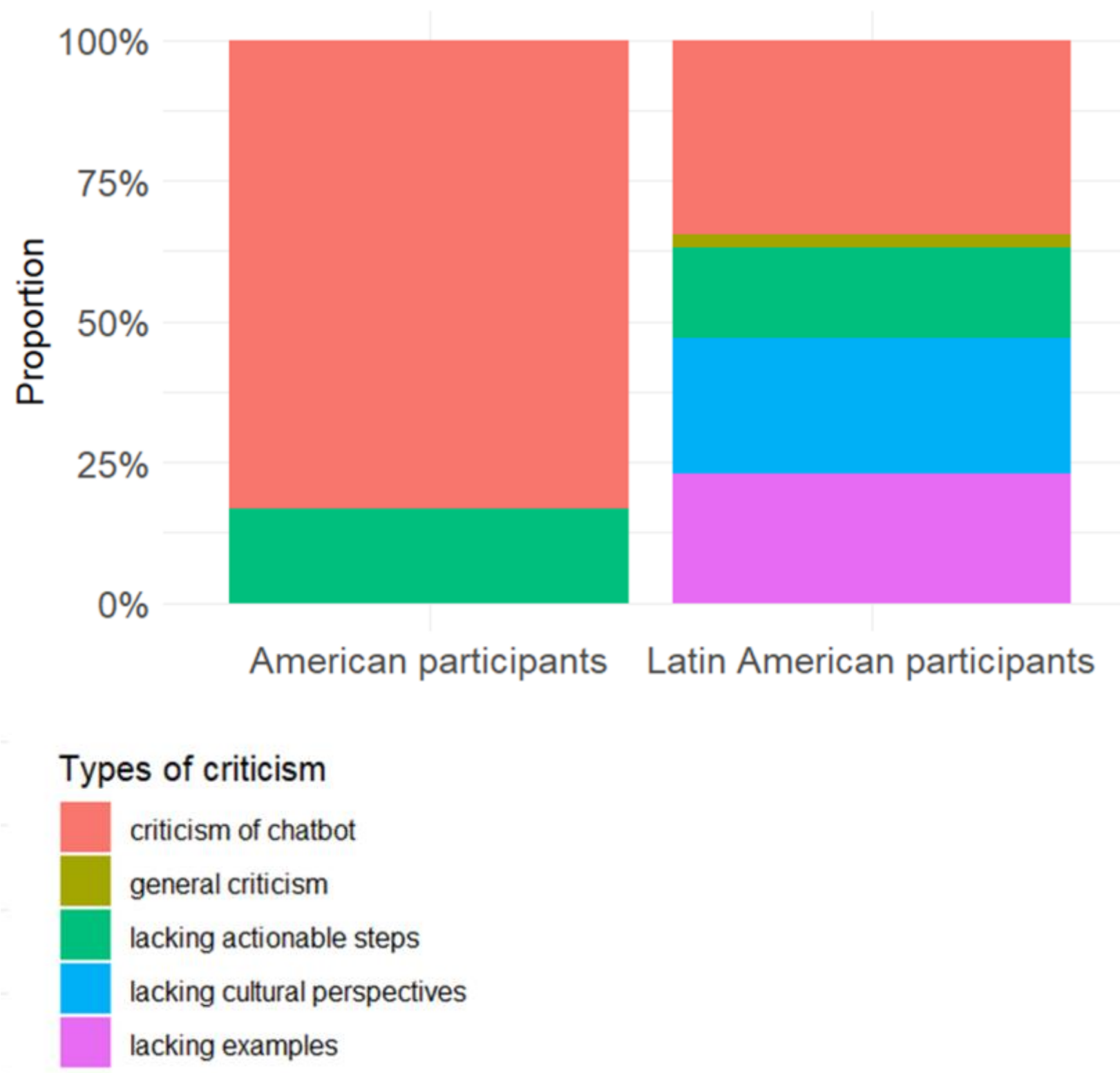

**Figure 4. What U.S. and Latin American participants in the own-cultural deliberation condition wrote when they gave a negative tag to a chatbot's response during the feedback session**

**Table 1. Results of mediation analysis.**

| | | | 95% CI | |
|---|---|---|---|---|
| **Indirect effects** | ***β*** | ***SE*** | **LL** | **UL** |
| Culturally not aligned deliberator→ positive emotion → intercultural empathy | 0.020 | 0.009 | 0.002 | 0.039 |
| American sample only | 0.038 | 0.018 | 0.003 | 0.073 |
| Latin American sample only | 0.013 | 0.012 | -0.010 | 0.036 |
| Culturally not aligned deliberator → bot satisfaction → intercultural empathy | 0.028 | 0.011 | 0.006 | 0.050 |
| American sample only | 0.053 | 0.021 | 0.012 | 0.094 |
| Latin American sample only | -0.001 | 0.010 | -0.020 | 0.018 |
| Culturally not aligned deliberator→ positive emotion → perspective-taking | 0.008 | 0.011 | -0.013 | 0.028 |
| American sample only | 0.013 | 0.021 | -0.028 | 0.054 |
| Latin American sample only | 0.012 | 0.016 | -0.019 | 0.043 |
| Culturally not aligned deliberator→ bot satisfaction → perspective-taking | 0.036 | 0.016 | 0.004 | 0.068 |
| American sample only | 0.065 | 0.030 | 0.005 | 0.125 |
| Latin American sample only | -0.002 | 0.018 | -0.038 | 0.034 |

*Note*. CI = confidence interval; LL = lower limit; UL = upper limit